\documentclass[prb,twocolumn,showpacs]{revtex4}

\usepackage{graphicx}
\begin{document}

\title{
   Fractional Quantum Hall States of Clustered Composite Fermions}

\author{
   Arkadiusz W\'ojs$^{1,2}$, 
   Kyung-Soo Yi$^{1,3}$, and 
   John J. Quinn$^1$}

\affiliation{
   $^1$University of Tennessee, Knoxville, Tennessee 37996, USA\\
   $^2$Wroclaw University of Technology, 50-370 Wroclaw, Poland\\
   $^3$Pusan National University, Pusan 609-735, Korea}

\begin{abstract}
The energy spectra and wavefunctions of up to 14 interacting 
quasielectrons (QE's) in the Laughlin $\nu={1\over3}$ fractional 
quantum Hall (FQH) state are investigated using exact numerical 
diagonalization.
It is shown that at sufficiently high density the QE's form pairs
or larger clusters.
This behavior, opposite to Laughlin correlations, invalidates the
(sometimes invoked) reapplication of the composite fermion picture 
to the individual QE's.
The series of finite-size incompressible ground states are 
identified at the QE filling factors $\nu_{\rm QE}={1\over2}$, 
${1\over3}$, ${2\over3}$, corresponding to the electron 
fillings $\nu={3\over8}$, ${4\over11}$, ${5\over13}$.
The equivalent quasihole (QH) states occur at $\nu_{\rm QH}=
{1\over4}$, ${1\over5}$, ${2\over7}$, corresponding to $\nu=
{3\over10}$, ${4\over13}$, ${5\over17}$.
All these six novel FQH states were recently discovered 
experimentally.
Detailed analysis indicates that QE or QH correlations in these 
states are different from those of well-known FQH electron states 
(e.g., Laughlin or Moore--Read states), leaving the origin of 
their incompressibility uncertain.
Halperin's idea of Laughlin states of QP pairs is also explored, 
but is does not seem adequate.
\end{abstract}
\maketitle

\section{Introduction}

Pan {\sl et al.}\cite{pan} recently (and Goldman and Shayegan
\cite{goldman} a little earlier) observed fractional quantum Hall 
(FQH) effect\cite{tsui,laughlin} in a two-dimensional electron 
gas at novel filling fractions $\nu$ of the lowest Landau level (LL$_0$).
The new FQH states are found to be spin-polarized and occur between 
the neighboring $\nu={1\over3}$ and ${2\over5}$ states of the Jain 
sequence,\cite{jain} corresponding to one and two completely filled 
composite fermion (CF) LL's, respectively.
Some of them, such as $\nu={4\over11}$ or ${4\over13}$,\cite{goldman} 
appear in the standard hierarchy of quasiparticle (QP) condensates 
formulated by Haldane\cite{haldane83} and Halperin,\cite{halperin84} 
but their ``hierarchical'' interpretation was earlier ruled out
\cite{hierarchy} because of the specific (subharmonic) form of the 
QP--QP interaction.
Others, such as the $\nu={3\over8}$ or ${3\over10}$ states, 
do not belong to the Haldane hierarchy, and the origin of their 
incompressibility is puzzling in an even more obvious way.

Recently there have been attempts\cite{smet} to explain these 
states in terms of ``multi-flavor'' CF pictures, with coexisting 
CF's carrying different numbers of fluxes (vortices).
Being equivalent to the CF hierarchy\cite{sitko-cfh} of 
Laughlin-correlated spin-polarized QP states, not only are these 
models not conceptually new, but they are also unjustified at the 
QP fillings in question.\cite{hierarchy}
In fact, some of the numerical results obtained earlier in a slightly 
different context\cite{parentage,fivehalf} and more detailed results 
presented here show that they {\em do not} describe the new FQH states. 

The appeal of the CF models lies in the fact that they allow one 
to think of a more complicated FQH state at filling factor $\nu$ 
as a CF-analogue of a simpler and better understood state 
at an effective CF filling factor $\nu^*$.
The best known examples are $\nu={1\over3}$ and ${2\over5}$ states 
corresponding to $\nu^*=1$ and 2, respectively.
In the present case $\nu={4\over11}$ and ${3\over8}$ correspond to 
$\nu^*={4\over3}$ and ${3\over2}$, that is to one-third and one-half 
filled first excited CF LL (CF-LL$_1$), respectively.
However, it cannot come as a surprise that the CF model does not
always work.
It is valid only for systems which support Laughlin correlations, 
and those only occur if the interactions are sufficiently strong 
at short range.\cite{haldane85}
Moreover, it is known precisely that ``sufficiently strong'' means 
that the interaction pseudopotential is superharmonic.
\cite{parentage,fivehalf}
It should also be noted that the CF analogy is not sufficient to 
explain all new observed states regardless of the fact that the 
(reapplied) CF model {\em does not}\cite{hierarchy} generally 
describe correlations between the CF's themselves.
Specifically, the $\nu={3\over10}$ state corresponds to $\nu^*=
{3\over4}$, while the electronic state at this filling is {\em not} 
incompressible.

The aim of this paper is to present the results of our ``numerical 
experiments'' for the new FQH states and show that they are described
by grouping of QP's into pairs\cite{qepair,flohr} (although probably 
without pair--pair Laughlin correlations\cite{halperin83}) or 
into larger clusters,\cite{read} rather than by a multi-flavor CF 
picture.
To do so, we: 
(i) present results of extensive numerical diagonalization studies 
of up to 14 interacting Laughlin quasielectrons (QE's); 
(ii) demonstrate directly the QE clustering by the calculation of 
pair amplitudes;
(iii) identify the series of finite-size nondegenerate ground states 
at $\nu_{\rm QE}={1\over2}$, ${1\over3}$, ${2\over3}$, corresponding 
to $\nu={3\over8}$, ${4\over11}$, ${5\over13}$; 
(iv) investigate the dependence of the stability of these states 
on the details of the QE--QE interaction pseudopotential;
(v) calculate their pair-correlation functions;
(vi) show that they have different QE--QE correlations than those 
of electrons in the Laughlin\cite{laughlin} or Moore--Read
\cite{moore,greiter,morf} states; 
(vii) construct the equivalent quasihole (QH) states at $\nu_{\rm QH}
={1\over4}$, ${1\over5}$, ${2\over7}$, corresponding to $\nu=
{3\over10}$, ${4\over13}$, ${5\over17}$; 
(viii) discuss a recent model\cite{qepair} assuming pairing of 
QP's and Laughlin correlations between the pairs (originally 
proposed by Halperin\cite{halperin83} for electrons in LL$_0$); 
and (ix) propose an explanation for the FQH state observed at 
$\nu={6\over17}$.

Standard numerical calculations for $N_e$ electrons are not useful 
for studying the new states, because convincing results require too 
large values of $N_e$.
Therefore, we take advantage of the knowledge
\cite{hierarchy,sitko-fl,lee01,lee02} of the dominant features of the 
pseudopotential $V_{\rm QE}(\mathcal{R})$ of the QE--QE interaction 
(i.e., the QE--QE interaction energy $V_{\rm QE}$ as a function of 
relative pair angular momentum $\mathcal{R}$), and diagonalize the 
(much smaller) interaction Hamiltonian of the $N$-QE systems.
This procedure was earlier shown\cite{sitko-fl} to accurately 
reproduce the low-energy $N_e$-electron spectra at filling factors
$\nu$ between ${1\over3}$ and ${2\over5}$.
It was also used in a similar, many-QE calculation by Lee {\sl et al.}
\cite{lee02} (who, however, have not found support for QE clustering). 

Our results confirm series of nondegenerate ground states with 
finite excitation gaps at $2l$, twice the QE angular momentum, 
equal to $2N-3$, $3N-7$, and ${3\over2}N+2$ (the last two states 
are particle--hole symmetric).
These series extrapolate to $\nu_{\rm QE}\equiv N/(2l+1)={1\over2}$, 
${1\over3}$, and ${2\over3}$, and to the electron filling factors 
$\nu={3\over8}$, ${4\over11}$, and ${5\over13}$, respectively.
The fact that the $\nu_{\rm QE}={1\over3}$ sequence occurs at $2l=3N-7$ 
rather than $3N-3$ implies that this state is {\em not} a Laughlin
state of QE's (or CF's).
Indeed, the assumption that the $\nu_{\rm QE}={1\over3}$ sequence 
must be described by the relation $2l=3N-3$ led to its being overlooked
in earlier finite-size calculations.\cite{mandal}
The identified sequence is also different from $2l=3N-5$ characteristic
of Halperin's paired state\cite{halperin83} corresponding to $\nu_{\rm 
QE}={1\over3}$.
On the other hand, the value of $2l=2N-3$ for the $\nu_{\rm QE}=
{1\over2}$ sequence suggests that this state could be a Halperin paired 
QE state (Laughlin state of QE pairs)\cite{halperin83,fivehalf,qepair} 
similar to the Moore--Read\cite{moore,greiter,morf} state of electrons 
at the half-filling of LL$_1$.
However, the squared overlaps with the Moore--Read state are very small 
($\sim0.03$ for $N\le14$), and the nondegenerate ground states occur in 
this series only for odd numbers of QE pairs (${1\over2}N=3$, 5, and 7),
which implies that the nature of this state is different.

The comparison of the QH--QH and QE--QE pseudopotentials (which 
differ mainly by a hard-core at $\mathcal{R}=1$ for the QH's) 
result in the following correspondence relation for the 
incompressible QH and QE states
\begin{equation}
\label{eq_qeqh}
   \nu_{\rm QH}^{-1}=2+\nu_{\rm QE}^{-1}.
\end{equation}
For $\nu_{\rm QE}={1\over2}$, ${1\over3}$, and ${2\over3}$, 
this relation gives $\nu_{\rm QH}={1\over4}$, ${1\over5}$, and 
${2\over7}$, corresponding to $\nu={3\over10}$, ${4\over13}$, 
and ${5\over17}$, respectively, all of which have also been 
observed experimentally.

To understand the origin of incompressibility in the new states
we explore an idea\cite{fivehalf,qepair} of Laughlin states of 
QP pairs (originally proposed by Halperin\cite{halperin83} to 
describe electron pairing in LL$_0$).
Grouping of QE's or QH's into pairs or even larger clusters at 
sufficiently large filling factors can be predicted from the form 
of QE--QE and QH--QH pseudopotentials, characterized by strong 
minima at $\mathcal{R}_{\rm QE}=1$ state and $\mathcal{R}_{\rm QH}=3$.
It is clearly demonstrated by the calculation of the appropriate 
pair amplitude coefficients\cite{haldane87} (related to the 
fractional grandparentage\cite{cowan}) in the many-QE ground 
states.
In Halperin's paired state, Laughlin correlations between the QP 
pairs would depend on the superharmonic behavior of the pair--pair 
interaction pseudopotential, $V_{{\rm QP}_2}(\mathcal{R}_2)$, at the 
relevant values of $\mathcal{R}_2$, the relative angular momentum 
of two pairs.
The analysis of the calculated $V_{{\rm QE}_2}(\mathcal{R}_2)$ 
suggests that of the whole sequence of incompressible Laughlin 
states of QE pairs, only $\nu_{\rm QE}={1\over2}$ might satisfy 
the condition for Laughlin correlations.
This appears to be in agreement with our ``numerical experiments,'' 
which reveal the $\nu_{\rm QE}={1\over2}$ series at $2l=2N-3$ 
(as predicted for Halperin's paired state) and the $\nu_{\rm QE}
={1\over3}$ series at $2l=3N-7$ (different from $3N-5$ of a 
Halperin's paired state).
However, as mentioned above, we find several strong indications 
that Halperin's paired state does not occur for QE's at neither
$\nu_{\rm QE}={1\over2}$ nor ${1\over3}$.

\section{Pseudopotentials, Laughlin Correlations, 
         and the Composite Fermion Picture}

The essential information about the interaction of particles
confined to some Hilbert space can be obtained by defining
the value of interaction energy for all allowed pair states.
For charged particles confined to a LL in the presence of 
a magnetic field, the relative motion is strongly quantized.
The orbital pair eigenstates can be labeled with a single discrete 
quantum number, relative angular momentum $\mathcal{R}$.
This number is a non-negative integer; it must be odd (even) for 
a pair of identical fermions (bosons), and it increases with
increasing average distance $\sqrt{\left<r^2\right>}$ between 
the two particles.

In Haldane's spherical geometry,\cite{haldane83} most convenient 
for finite-size calculations, the LL$_0$ is represented by 
a degenerate shell of single-particle angular momentum $l=Q$.
Here $2Q\phi_0$ is the strength of Dirac monopole in the center, 
defined as $4\pi R^2 B$, the total flux of the magnetic field $B$ 
through the surface of radius $R$ (using the definition of the 
magnetic length $\lambda=\sqrt{\hbar c/eB}$, this can be written 
as $Q\lambda^2=R^2$).
The total pair angular momentum $L'$ (here, $L$ means total angular 
momentum of $N$ particles, and $L'$ is reserved for $N=2$) results 
from an addition of two angular momenta $l$ of individual particles, 
and it is connected to the relative pair angular momentum via relation 
$L'=2l-\mathcal{R}$.
Thus, the maximum value of $L'=2l$ (for bosons) or $2l-1$ (for fermions) 
corresponds to the smallest pair state with $\mathcal{R}=0$ or 1.

The pair interaction energy $V$ expressed as a function of 
$\mathcal{R}$ is called the pseudopotential, and the series of 
its parameters $V(\mathcal{R})$ entirely determines many-body 
correlations.
On a sphere, $\mathcal{R}\le2l$ and thus the number of pseudopotential 
parameters is finite.
However, even in an infinite (planar) system, only those few leading 
parameters at the values of $\mathcal{R}$ corresponding to the average 
distance $\sqrt{\left<r^2\right>}$ not exceeding the correlation length 
$\xi$ are of significance (provided that the correlations are indeed 
characterized by finite $\xi\sim\lambda$).

Remarkably, even for the completely repulsive interactions, different 
correlations can result in a partially filled shell depending on the 
form of $V(\mathcal{R})$. 
For example, if $V$ {\em increases} as a function of $\mathcal{R}$ 
(as in atomic shells in the absence of magnetic field), the low-energy 
many-body states obey Hund's rule and tend to have the maximum 
possible degeneracy (i.e., the maximum $2L+1$).
In the opposite extreme situation, when $V$ {\em decreases} 
sufficiently quickly\cite{haldane85} as a function of $\mathcal{R}$, 
Laughlin correlations occur.
These correlations are defined as the tendency to avoid pair states 
with one or more smallest values of $\mathcal{R}$, i.e., with the 
largest repulsion (the relative occupation of different pair 
states in a many-body state is a well-defined quantity, given by 
the pair amplitude coefficient\cite{haldane87}).

As a result of Laughlin correlations, the low-energy many-body 
states usually have small degeneracy and effects commonly 
associated with the FQH physics occur, including the formation 
of incompressible ground states at certain values of $\nu$.
What is often not realized or overlooked is that it is precisely 
the Laughlin correlations that justify the CF picture.
In other words, the mean field CF picture that attaches $2p$ 
magnetic flux quanta (or vortices) to each fermion and predicts 
the family of Jain wavefunctions for the lowest energy states 
is correct {\em if and only if} those fermions have Laughlin 
correlations, i.e., the lowest energy states indeed maximally 
avoid having pair states with $\mathcal{R}\le(2p+1)$.
For example, in order to bind $2p$ vortices and transform into 
CF's, electrons must have Laughlin correlations (and indeed they 
do in LL$_0$).
These CF's (or, more precisely, the QP's in partially filled 
CF-LL's) would bind additional vortices and turn into 
``higher-order'' CF's if they themselves had Laughlin correlations 
(and in this paper we show that, at the relevant filling factors, 
they {\em do not}).

Another important class of pseudopotentials are the ``harmonic'' 
ones, i.e., those for which parameters $V_{\rm H}(\mathcal{R})$ fall 
on a straight line when plotted as a function of the average squared 
distance $\left<r^2\right>$.
Clearly, all harmonic potentials $V_{\rm H}(r)=a_0+a_2r^2$ have this 
property regardless of the LL confinement.
It has been shown\cite{parentage} that for particles confined in 
an angular momentum shell on a sphere, $V_{\rm H}$ is a linear 
(increasing in case of repulsion) function of squared pair angular 
momentum, $L'(L'+1)$.
It follows from considering the large-radius limit ($R\rightarrow
\infty$ and $\lambda=const$) that on a plane (or on a ``large'' 
sphere, i.e., for $\mathcal{R}\ll2l$), $V_{\rm H}$ is a linear 
function of $\mathcal{R}$.
The importance of the harmonic pseudopotential lies in the fact that
it causes {\em no} correlations, i.e., all many-body states with the 
same total angular momentum $L$ are degenerate (and their energy is 
just a linear function of $L(L+1)$ or $\mathcal{R}$, depending on 
geometry).\cite{parentage}
It is thus only the anharmonic part of $V(\mathcal{R})$ that causes
correlations, while the harmonic part only shifts the entire energy 
spectrum by a constant times $L(L+1)$ or $\mathcal{R}$.

From the analysis of the sum rules\cite{sum-rules} obeyed by the pair 
amplitudes $\mathcal{G}_\Psi(\mathcal{R})$ measuring the fraction of 
pairs with relative pair angular momentum $\mathcal{R}$ out of the 
total number of ${1\over2}N(N-1)$ pairs in an $N$-particle state $\Psi$, 
it has been shown\cite{parentage,fivehalf} that Laughlin correlations 
occur near filling factor $\nu$ if the dominant anharmonic contribution 
to $V$ is positive at the avoided values of $\mathcal{R}$.
For example, for fermions at $\nu\approx(2p+1)^{-1}={1\over3}$, 
the pseudopotential $V(\mathcal{R})$ must decrease ``superlinearly'' 
through any three values $a<b<c$ beginning with $a=1$.
By the superlinear (i.e., superharmonic) behavior we mean that
\begin{equation}
\label{eqsh}
   {V(a)-V(b)\over b-a}>{V(b)-V(c)\over c-b}.
\end{equation}
Only then do Laughlin correlations occur and justify the use of the 
mean field CF transformation that attaches $2p=2$ fluxes (vortices) 
to each electron.
Moreover, any pseudopotential that is strongly superharmonic at
short range causes the same (Laughlin) correlations which explains
the robust character of the FQH states in realistic systems or
in model calculations.

It has been shown\cite{parentage} that it is the superharmonic 
behavior of the Coulomb repulsion $V(r)\sim r^{-1}$ in LL$_0$ 
in the entire range of $\mathcal{R}$ that explains the success 
of the CF picture through the entire Jain sequence of fractions.
\cite{jain}
It was also shown\cite{fivehalf} (by direct calculation of pair 
amplitudes) that because the Coulomb pseudopotential in LL$_1$ 
is roughly {\em linear} between $\mathcal{R}=1$ and 5, the 
electrons tend to form pairs with $\mathcal{R}=1$ when filling 
a fraction ${1\over4}\le\nu_1\le{1\over2}$ of LL$_1$.
This is exactly the opposite behavior to the avoidance of this 
pair state that would characterize a state with Laughlin 
correlations and that could justify the CF picture (sometimes 
erroneously used in literature to describe the FQH states 
at $\nu\equiv2+\nu_1={5\over2}$, ${7\over3}$, or ${8\over3}$).

Let us stress here that the mean field CF picture simply mimics
the fact that (in a Laughlin-correlated system) each electron 
drags a $\mathcal{R}=1$ correlation hole with it -- by replacing 
the ``bare'' electron LL degeneracy with an appropriately smaller, 
``effective'' one (and an effective CF magnetic field $B^*$ is just 
an intuitive physical picture that cannot be treated literally).
In fact, it has recently been demonstrated\cite{adiabatic} that 
the adiabatic addition of flux (instead of addition via gauge 
transformation) automatically gives rise to Laughlin correlations 
without the need of any mean field approximation.
Having said this, there simply are {\em no} CF's in the $\nu={5\over2}$,
${7\over3}$, or ${8\over3}$ states, let alone the CF pairs.
Instead, at least the Moore--Read state at $\nu={5\over2}$ is clearly
a paired state of {\em electrons}\cite{fivehalf,morf} (although
models involving pairing of CF's in this state can also be found 
in literature\cite{morinari}).
It is surprisingly often overlooked that the FQH effect does not
prove the existence of CF's or Laughlin correlations, but only 
the existence of a nondegenerate ground state separated from the 
continuum of QP excitations by a finite gap -- the property which 
can also result from correlations of a different nature.

It is indeed quite remarkable that the knowledge of the interaction 
pseudopotential $V(\mathcal{R})$ at short range is sufficient to 
predict or rule out Laughlin correlations in different FQH systems.
\cite{hierarchy,parentage,x-,skyrmions}
However, it must be carefully noticed that the predicted absence 
of Laughlin correlations does not preclude the FQH effect itself,
only a microscopic origin of the effect attributable to Laughlin
correlations.
It should also be realized that immediate application of the CF 
model without studying the interactions between the relevant 
particles (electrons, QP's, etc.) whenever real or numerical 
experiments reveal incompressibility is not justified.
Precisely such a situation was recently encountered with the 
discovery of new FQH states at $\nu={3\over8}$, ${4\over11}$, etc., 
which turn out not to be Laughlin or Jain states (of QP's) despite 
being incompressible.

\section{QP Interactions}

It follows from the preceding discussion that in order to explain 
the origin of incompressibility in the new FQH states, one has to 
begin with the identification of the relevant (quasi) particles 
(electrons, holes, Laughlin QP's, CF's, excitons, skyrmions, \dots), 
analyze their interaction pseudopotentials, understand their 
correlations, and finally derive the filling factors $\nu$ at 
which those correlations cause incompressibility.
In contrast to the CF model (which, nevertheless, is still very 
elegant and useful {\em after} it is proven valid for a particular
system), this line of thought is free of unproven assumptions, 
such as that of a cancellation between the Coulomb and gauge 
interactions beyond the mean field.

It is well-established that a (Laughlin-correlated) system of 
electrons at ${1\over3}\le\nu\le{2\over5}$ can be viewed as one 
of (fractionally charged and thus less strongly interacting) QE's 
moving in the underlying Laughlin $\nu={1\over3}$ ground state.
This is elegantly pictured in the CF model, in which the Laughlin 
state corresponds to the completely filled CF-LL$_0$, and the 
QE's correspond to the (weakly interacting) particles moving in 
the (partially filled) CF-LL$_1$.
Similarly, the electron system at ${1\over4}\le\nu\le{1\over3}$ 
can be viewed as the QH's moving over the $\nu={1\over3}$ 
background (with the QH's pictured as vacancies in CF-LL$_0$).

\begin{figure}
\resizebox{3.4in}{3.08in}{\includegraphics{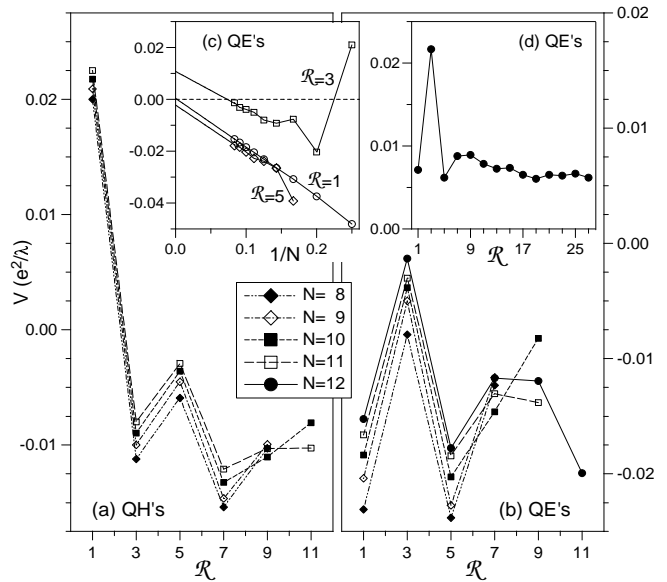}}
\caption{
\label{fig1}
   Interaction pseudopotentials $V(\mathcal{R})$ for the QH's (a) 
   and QE's (b) of the Laughlin $\nu={1\over3}$ state calculated 
   in the systems of up to $N=12$ electrons on a sphere.
   Insets:
   (c) Dependence of the leading QE--QE pseudopotential coefficients 
   corresponding to the smallest values of $\mathcal{R}$ on $N^{-1}$.
   Extrapolation to $N^{-1}\rightarrow0$ corresponds to an infinite 
   planar system.
   (d) QE--QE pseudopotential calculated by Lee {\sl et al.}
   \cite{lee01}} 
\end{figure}

Therefore, we begin the study of the new FQH states in the 
${1\over4}\le\nu\le{2\over5}$ range with the analysis of the 
QE--QE and QH--QH pseudopotentials.
In the following we will use the fermionic statistics to describe
QP's which is consistent with the CF picture (and conversion to 
bosons or anyons can be done in a standard way\cite{wilczek}).
The qualitative behavior of $V_{\rm QP}(\mathcal{R})$ at short 
range is well-known from the numerical studies of small systems.
\cite{hierarchy,sitko-fl,lee01,lee02}  
In Fig.~\ref{fig1}(b) we compare $V_{\rm QE}(\mathcal{R})$ 
calculated for the systems of $N=8$ to 12 electrons.
As the calculation involves subtraction of the $N$-electron energies 
corresponding to zero, one, and two QE's that (in finite systems) 
occur at different values of $Q=\sqrt{R/\lambda}$ (i.e., different 
surface curvatures), the zero of energy is determined much less
accurately than the relative values of different pseudopotential 
parameters.
However, when the data for each $\mathcal{R}$ are extrapolated to 
large $N$, the positive sign of $V_{\rm QE}(\mathcal{R})$ is restored, 
as shown in Fig.~\ref{fig1}(c).
Still, only the relative values are of importance, since adding 
a constant to $V(\mathcal{R})$ does not affect correlations and 
only shifts the whole many-body spectrum by a (different) constant.
On the other hand, the repulsive character of the QP--QP interaction
and the long-range behavior of $V_{\rm QP}(\mathcal{R})\sim
\mathcal{R}^{-1/2}$ follow from the fact that QP's are charged 
particles (the form of QP charge density affects $V_{\rm QP}$ 
only at short range, comparable to the QP size).
In Fig.~\ref{fig1}(d) we plot $V_{\rm QE}(\mathcal{R})$ obtained 
more recently by Lee {\sl et al.}\cite{lee01} using a somewhat 
different approach.
Since it confirms the oscillatory behavior at short range in 
Fig.~\ref{fig1}(b) and behaves as expected at long range, we 
will use it later to diagonalize interaction in the systems 
of more than two QE's.

Clearly, the dominant features of $V_{\rm QE}$ are the small value 
at $\mathcal{R}=1$ and a strong maximum at $\mathcal{R}=3$.
Similar analysis for $V_{\rm QH}(\mathcal{R})$ shown in 
Fig.~\ref{fig1}(a) for $8\le N\le11$ reveals the maxima at 
$\mathcal{R}=1$ and 5, and the nearly vanishing $V_{\rm QH}(3)$.
Actually, it follows from the comparison of Figs.~\ref{fig1}(a) 
and (b) that the slightly reduced energy scale for $V_{\rm QH}
(\mathcal{R})$ and the additional strongly repulsive state at 
$\mathcal{R}_{\rm QH}=1$ are the only significant differences 
between the two pseudopotentials.
The $V_{\rm QE}(\mathcal{R})\sim V_{\rm QH}(\mathcal{R}+2)$
correspondence and the fact that $V_{\rm QH}(1)$ is the largest 
of all QE or QH parameters will be used to construct the QH 
states corresponding to the incompressible QE states studied 
numerically in detail.

The above conclusions about the properties of QP--QP pseudopotentials 
weakly depend on such assumptions as zero layer thickness $w$ 
or infinite magnetic field $B$, and their oscillations at small 
$\mathcal{R}$ persist in realistic FQH systems.
It is noteworthy that this result cannot be obtained from the 
{\em literally understood} original formulation of the CF model 
in which the weak ``residual'' CF--CF interactions are said to 
result from partial cancellation of strong Coulomb and gauge 
interactions between the electrons.
This is because these two interactions have different character 
and, for example, depend differently on $w$ or $B$.\cite{parentage}

\section{Corresponding QE and QH States}

It can be seen in Fig.~\ref{fig1} that $V_{\rm QH}(1)$ is the 
strongest anharmonic contribution to $V_{\rm QH}(\mathcal{R})$.
This causes the maximum avoidance of the two-QH state with 
$\mathcal{R}=1$ (Laughlin QH--QH correlations) and justifies 
the CF transformation with $2p=2$ fluxes attached to each QH 
(i.e., such reapplication of the CF transformation to the 
vacancies in the partially filled CF-LL$_0$).
The states of CF-QH's obtained in this way form the lowest band
of QH states at their filling factors $\nu_{\rm QH}\le{1\over3}$.
At $\nu_{\rm QH}={1\over3}$, the QH Laughlin state occurs
that corresponds to the $\nu={2\over7}$ hierarchy/Jain state.
At $\nu_{\rm QH}={1\over5}$, the CF-QH's (unlike electrons) do 
not bind any more vortices because of the subharmonic character 
of $V_{\rm QH}(\mathcal{R})$ around $\mathcal{R}=3$.

If follows from the $V_{\rm QE}(\mathcal{R})\sim V_{\rm QH}
(\mathcal{R}+2)$ correspondence seen in Fig.~\ref{fig1} that the 
pseudopotential for the interacting CF-QH's is similar to that 
of QE's.
To confirm this, we have calculated this pseudopotential in a
standard way,\cite{hierarchy} by numerical diagonalization of 
$N$ QH's interacting through $V_{\rm QH}$ in a shell of angular 
momentum $l_{\rm QH}={3\over2}(N-1)+2$.
The similarity between $V_{\rm QE}(\mathcal{R})$ and 
$V_{\rm CF-QH}(\mathcal{R})$ not only confirms that no additional 
fluxes can be attached to the CF-QH's (i.e., not more than two 
fluxes to the original QH's), but it also implies that the same 
correlations will occur in QE and CF-QH systems, and that any 
incompressible QE state must have its CF-QH counterpart at the 
same filling factor.

The conversion of the CF-QH filling factors to $\nu_{\rm QH}$ 
gives Eq.~(\ref{eq_qeqh}), connecting the observed states into 
pairs: $\nu=$ (${3\over8}$ and ${3\over10}$), (${4\over11}$ 
and ${4\over13}$), and (${5\over13}$ and ${5\over17}$).
Using the hierarchy equation, $\nu^{-1}=2+(\nu^*)^{-1}$ where 
$\nu^*=1+\nu_{\rm QE}$ or $1-\nu_{\rm QH}$, it can also easily 
be shown that the two fractions in each pair $(\nu,\mu)$ 
are connected by
\begin{equation}
   \nu^{-1}+\mu^{-1}=6.
\end{equation}

\section{Finite-Width Effect at $\nu={6\over17}$}

While the hierarchy interpretation is certainly invalid for the
three pairs of states discussed in the preceding section, and 
an alternative explanation must exist for their incompressibility, 
the situation with another observed state, $\nu={6\over17}$, 
corresponding to $\nu_{\rm QE}={1\over5}$, is less obvious.
Its QH counterpart at $\nu={6\over19}$ has not been observed, 
and it is not clear if the finite width $w$ of the actual 
experimental system (which tends to weaken oscillations in 
$V_{\rm QP}$) does not lift $V_{\rm QE}(1)$ enough compared 
to $V_{\rm QE}(\mathcal{R}\ge5)$ that avoiding both 
$\mathcal{R}=1$ and 3 at the same time (i.e., formation of 
the Laughlin state of the QE's with $\nu_{\rm QE}={1\over5}$ 
as assumed in the CF hierarchy picture) becomes energetically 
favorable.
If true, this would be a similar scenario to that in LL$_1$, 
where the $\nu={1\over3}$ state {\em is not} a Laughlin state, 
but the $\nu={1\over5}$ state {\em is}. 
If the $\nu={6\over17}$ state could indeed only be observed
in sufficiently wide electron systems, then it is possible 
that the unobserved $\nu={6\over19}$ state (corresponding to 
$\nu_{\rm QE}={1\over7}$) would simply require slightly larger 
width to become incompressible.

The difference between critical widths could probably be explained 
by the fact that QH--QH pseudopotential parameter that must be
lifted is at a larger $\mathcal{R}$ (at 3 instead of 1) which 
thus corresponds to a larger average in-plane QH--QH separation
$\sqrt{\left<r^2\right>}$.
Unfortunately, our estimates of the $V_{\rm QP}(\mathcal{R})$ 
pseudopotentials are not sufficiently accurate to make definite 
predictions about the critical widths.
However, the $N=10$ electron calculation for the QE's shows that 
$V_{\rm QE}(1)$ indeed moves up relative to $V_{\rm QE}(5)$ and
$V_{\rm QE}(7)$ when the width is increased from $w=0$ to 20~nm.
Similar behavior was found for $V_{\rm QH}$ calculated for
$N=8$: the $V_{\rm QH}(3)$ moved up relative to $V_{\rm QH}(7)$ 
and $V_{\rm QH}(9)$ with increasing width, only at a smaller 
rate $dV/dw$ than it did for QE's.

\section{QP Clustering}

Although in the following discussion of QP states we will concentrate 
on the QE's, the extension to QH's remains valid as discussed above.
Even without further numerical proof it is evident from 
Fig.~\ref{fig1} alone that the QE's interacting through $V_{\rm QE}
(\mathcal{R})$ will not have Laughlin correlations.
This implies that the mean field CF transformation {\em cannot} be
reapplied to the particles or vacancies in CF-LL$_1$.
This rules out the simple hierarchy picture of the $\nu={4\over11}$ 
state, as well as the (equivalent though even less justified) 
interpretation involving the coexistence of CF's carrying two and 
four flux quanta (or vortices).\cite{smet}
In the latter, ``multi-flavor'' CF model, the CF's carrying two 
additional flux quanta are constructed by a reapplication of the 
CF transformation to those QP's in the ${1\over3}$-filled CF-LL$_1$.
This procedure was actually first proposed by Sitko {\sl et al.},
\cite{sitko-cfh} so it is not new, and it is equivalent to the Haldane 
hierarchy (except that it is expressed in terms of fermionic rather 
than bosonic QE's compared to Haldane's original paper\cite{haldane83}).
Furthermore, it has been clearly demonstrated\cite{parentage} in small 
systems with superharmonic pseudopotentials $V$ that adding $2p=2$, 4, 
\dots\ flux quanta to each particle in a mean field CF transformation 
partitions the entire many-body Hilbert space into subspaces separated 
by energy gaps associated with the avoided $V(\mathcal{R})$.

What are these non-Laughlin QE--QE correlations?
Clearly, the avoided pair state must now be $\mathcal{R}=3$ while
having pairs in the weakly repulsive $\mathcal{R}=1$ state does not
increase the total interaction energy $E$ given by
\begin{equation}
   E={1\over2}N(N-1)
     \sum_\mathcal{R}\mathcal{G}(\mathcal{R})V(\mathcal{R}),
\end{equation}
where $\mathcal{G}(\mathcal{R})$ denotes the pair amplitude 
(i.e., the fraction of pairs with relative pair angular momentum 
$\mathcal{R}$). 
Therefore at least some of the QE's will form such pairs (QE$_2$'s)
or even larger clusters (QE$_K$'s) at filling factors $\nu_{\rm QE}>
{1\over5}$ (when the avoidance of both $\mathcal{R}=1$ and 3 at the 
same time is not possible).
Let us stress that the proposed clustering is not a result of some 
attractive QE--QE interaction,\cite{qepair} but due to an obvious 
tendency to avoid the strongly repulsive $\mathcal{R}=3$ pair state 
in a system of sufficiently large density.

As an illustration for such clustering, consider a system 
of one-dimensional classical point charges moving along the $z$-axis, 
at a fixed linear density $dN/dz=1$, and interacting through 
a repulsive potential $V(z)$.
Let us compare the following two configurations: (a) equally spread 
particles at $z_k=k$, and (b) pairs at $z_{2k}=z_{2k+1}=k$, where 
$k=0$, $\pm1$,  $\pm2$, \dots.
The difference between the total energies counted per one particle 
is $\varepsilon_b-\varepsilon_a={1\over2}V(0)-\sum_{k=1}^\infty 
[V(2k-1)-V(2k)]$, and it can have either sign depending on the 
form of $V(z)$.
For example, if $V(z)=|z|^{-1}$ at $|z|\ge1$, then the paired 
configuration (b) has lower energy if $V(z)<2\ln 2$ at short 
range.
For such form of $V(z)$, the transition between configurations 
(a) and (b) will occur at sufficiently high density $dN/dz$.

\begin{figure}
\resizebox{3.4in}{1.93in}{\includegraphics{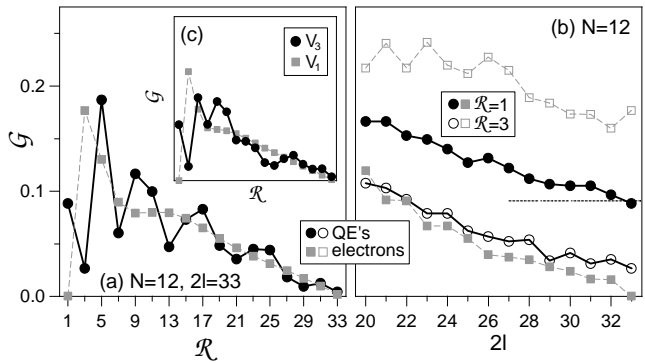}}
\caption{
\label{fig2}
   (a) 
   Pair-correlation functions (pair amplitude $\mathcal{G}$ 
   as a function of relative pair angular momentum $\mathcal{R}$) 
   for the lowest $L=0$ eigenstates of $N=12$ particles on Haldane 
   sphere with $2l=33$. 
   Gray squares are for the electrons in LL$_0$ (approximate Laughlin 
   $\nu={1\over3}$ state) and black dots are for the QE's interacting 
   through the pseudopotential of Fig.~\ref{fig1}(d).
   Inset (c) shows the same for model pseudopotentials $V_1$ and $V_3$.
   (b) 
   Dependence of the two leading pair amplitudes, $\mathcal{G}(1)$ 
   and $\mathcal{G}(3)$, on $2l$ for $N=12$ electrons (squares) and 
   QE's (dots).
   The horizontal dotted line indicates the value $\mathcal{G}=(N-1)^{-1}$
   corresponding to ${1\over2}N$ widely separated pairs.}
\end{figure}

A clustered state proposed here for the QE's would be characterized 
by a greatly reduced pair amplitude $\mathcal{G}(3)$ compared to 
the Laughlin $\nu={1\over3}$ state in order to minimize the total 
energy.
At the same time, the value of $\mathcal{G}(1)$ would be increased 
from nearly zero to a value of the order of $(N-1)^{-1}$ describing 
all $N$ QE's forming ${1\over2}N$ (relatively widely separated) pairs.
This behavior is demonstrated in Fig.~\ref{fig2}(a), in which we 
compare $\mathcal{G}$ plotted as a function of $\mathcal{R}$, 
calculated for the lowest states with total angular momentum $L=0$ 
in systems of $N=12$ particles in the shell with $2l=33$, interacting 
through different pseudopotentials.
The $\mathcal{G}(\mathcal{R})$ is a form of pair-correlation function, 
more suitable to analyze correlations in a Hilbert space restricted to 
LL$_0$ than the ``real-space'' pair-correlation function $\mathcal{G}(r)$.
It is defined\cite{haldane87} in terms of the projection operator 
$\mathcal{P}_{ij}(\mathcal{R})$ onto the subspace in which pair $\left<
ij\right>$ is in the eigenstate $\left|\mathcal{R}\right>$, and it can
readily be calculated\cite{parentage} (using eigenfunctions of the 
actual pseudopotential) as an expectation value of a ``selective'' 
interaction pseudopotential $V_\alpha(\mathcal{R})=\delta_{\alpha
\mathcal{R}}$,
\begin{equation}
   \mathcal{G}(\mathcal{R})=\left<V_\mathcal{R}\right>.
\end{equation}
The squares in Fig.~\ref{fig2}(a) correspond to the ground state of 
electrons interacting through the Coulomb potential in LL$_0$.
The full dots describe the QE's interacting through the pseudopotential 
shown to in Fig.~\ref{fig1}(d).
In the inset (c), the squares and circles describe the ground states 
of selective interactions $V_\alpha(\mathcal{R})$.
The ground state of $V_1$ is the exact Laughlin $\nu={1\over3}$ 
wavefunction, and $V_3$ remarkably well reproduces correlations 
of the QE system, which proves that it is the ability to avoid 
$\mathcal{R}=3$ that selects the low-energy many-QE states.
The significant reduction of $\mathcal{G}(3)$ and an increase of 
$\mathcal{G}(1)$ when going from electrons to QE's are also 
clearly visible.

Since the reason for the QE clustering is the avoidance of 
$\mathcal{R}=3$ rather than QE--QE attraction, it seems reasonable 
to assume that some of the clusters should break up at lower filling 
factors.
Mixed states of pairs and unpaired electrons have been proposed 
earlier in attempt to explain the $\nu={7\over3}$ state in LL$_1$,
\cite{fivehalf} but here we have not found evidence for such 
behavior down to $\nu_{\rm QE}={1\over3}$.
In Fig.~\ref{fig2}(b) we plot $\mathcal{G}(1)$ and $\mathcal{G}(3)$, 
calculated in the lowest $L=0$ states of $N=12$ particles (electrons
and QE's), as a function of $2l$.
For the QE's, as $2l$ increases from 20 to 33 (i.e., $\nu$ decreases 
from $\sim{1\over2}$ to $\sim{1\over3}$), the $\mathcal{G}(3)$ 
decreases to zero while $\mathcal{G}(1)$ remains larger than 
$(N-1)^{-1}$, the value corresponding to the widely separated 
${1\over2}N$ pairs.

\section{Interaction of QP Pairs}

If the QP fluid consisted of QP$_2$ molecules, the QP$_2$--QP$_2$ 
interactions would need to be studied to understand correlations.
Being pairs of fermions, the QP$_2$'s will be treated as bosons
carrying angular momentum $l_{{\rm QP}_2}^{\rm boson}\equiv 
l_{{\rm QP}_2}=2l-\mathcal{R}_{\rm QP}=2l-1$ for the QE$_2$'s 
and $2l-3$ for the QH$_2$'s.
However, in two dimensions they can be easily converted to fermions 
by a transformation consisting of attachment of one flux quantum,
\cite{wilczek} i.e., by an adjustment of angular momentum, 
$l_{{\rm QP}_2}^{\rm fermion}=l_{{\rm QP}_2}^{\rm boson}+{1\over2}
(N_2-1)$, where $N_2={1\over2}N$ is the number of pairs.
The QP$_2$--QP$_2$ interaction is described by an effective 
pseudopotential $V_{{\rm QP}_2}(\mathcal{R}_2)$ that includes 
correlation effects caused by the fact that the two-pair 
wavefunction must be symmetric under exchange of QP$_2$ bosons 
and at the same time antisymmetric under exchange of any two 
QP fermions.

In order to calculate this pseudopotential one must solve the 
problem of the stability of two QP$_2$'s in the absence of the
surrounding QP's.
We have done it by constructing trial paired wavefunctions
$\left|\mathcal{R}_2\right>_{\rm pair}$ in the following way.
The four QP's are divided into two pairs distinguished by two 
projections of pseudospin, $\sigma=\,\uparrow$ and $\downarrow$.
A $\sigma$-asymmetric pairing interaction is defined as 
$V_{\sigma\sigma'}(\mathcal{R})=-\delta_{\sigma\sigma'}
\delta_{\mathcal{R}\mathcal{R}_{\rm QP}}$ with $\mathcal{R}_{\rm 
QE}=1$ and $\mathcal{R}_{\rm QH}=3$.
It is diagonalized in the basis of totally antisymmetric four-QP 
states, i.e., in the subspace of maximum total pseudospin.
The resulting lowest-energy eigenstates at each angular momentum $L$
are the ``maximally paired'' states $\left|\mathcal{R}_2\right>_{\rm 
pair}$ corresponding to the relative angular momentum $\mathcal{R}_2
=2l_{{\rm QP}_2}-L$.
By ``maximally paired'' we mean here that these states have the 
largest possible pair amplitude 
$\mathcal{G}_{\uparrow\uparrow}(\mathcal{R}_{\rm QP})
+\mathcal{G}_{\downarrow\downarrow}(\mathcal{R}_{\rm QP})$ 
which is simply equal to the negative of the eigenvalue of the
pairing interaction energy.
The ``complete pairing'' corresponding to the eigenenergy equal 
to $-2$ is not allowed for identical QP's, i.e., in the subspace 
of maximum total pseudospin, because the three angular momenta, 
$\mathcal{R}_{\rm QP}$ for each pair and $\mathcal{R}_2$ describing 
relative motion of the two pairs, cannot be simultaneously conserved.

The relaxation of the angular momentum (and thus also of energy) 
of each of the two pairs that come in contact is due to the 
appropriate required symmetry of the total two-pair state with 
respect to an interchange of the individual QP's.
This is a statistics-induced correlation effect, independent of 
the electric interaction between the pairs (it also occurs for 
the model pairing interaction that vanishes for a pair of QP's 
that belong to different pairs).
The pair--pair pseudopotential $V_{{\rm QP}_2}(\mathcal{R}_2)$, 
calculated as the expectation energy of $V_{\rm QP}$ in the state 
$\left|\mathcal{R}_2\right>_{\rm pair}$, minus twice the energy 
of one pair, $2V_{\rm QP}(\mathcal{R}_{\rm QP})$, automatically 
includes this effect.
However, it must be realized that the pair--pair interaction 
is more complicated due to the internal structure of each pair 
that comes into play via statistics, and that at short range 
its description in terms of an effective pseudopotential is 
only an approximation.

\begin{figure}
\resizebox{3.4in}{1.87in}{\includegraphics{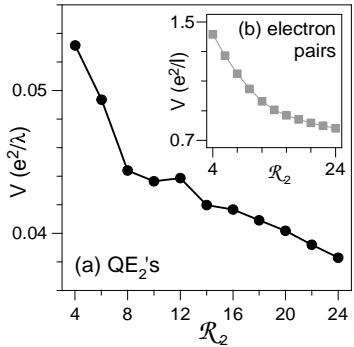}}
\caption{
\label{fig3}
   Short-range parts of the pseudopotentials $V(\mathcal{R}_2)$ 
   for the interaction between (a) two QE$_2$'s and (b) two 
   pairs of electrons in LL$_0$, calculated on Haldane sphere 
   for $2l=30$.}
\end{figure}

Fig.~\ref{fig3}(a) shows the result obtained for the QE's in 
a shell with $2l=30$, interacting through the pseudopotential 
of Fig.~\ref{fig1}(d).
The minimum value of $\mathcal{R}_2=4$ corresponds to the 
maximum-density four-particle droplet with $L=4l-6$, 
and we only show the data up to $\mathcal{R}_2=24$.
The $V_{{\rm QE}_2}(\mathcal{R}_2)$ appears weakly subharmonic 
at $\mathcal{R}_2=4$ (i.e., between $\mathcal{R}_2=4$ and 8 in 
the sense of definition (\ref{eqsh})), but it is clearly 
superharmonic at $\mathcal{R}_2=6$ (i.e., between $\mathcal{R}_2
=6$ and 10).
For comparison, in Fig.~\ref{fig3}(b) we show the superharmonic 
and rather featureless pair--pair pseudopotential for the 
electrons in LL$_0$.

\section{Halperin's Paired QP States}

If QP's formed pairs (QP$_2$'s) in a many-QP state, and if the 
pseudopotential $V_{{\rm QP}_2}(\mathcal{R}_2)$ were superharmonic 
in the entire range, then the QP$_2$'s would have Laughlin correlations.
Being bosons, they would then form a sequence of incompressible 
Laughlin states at $\nu=(2q)^{-1}$, characterized by having 
$\mathcal{R}_2\ge2q$ for all QP$_2$--QP$_2$ pairs.
These states have been originally proposed by Halperin\cite{halperin83} 
to describe such electron states in LL$_0$ as $\nu={2\over5}$.
Later they were often invoked\cite{greiter} in the context of FQH effect 
at $\nu={5\over2}$ to describe pairing of electrons in half-filled LL$_1$.
They can be conveniently described using the following ``composite 
boson'' (CB) model.\cite{fivehalf}

In spherical geometry, let us consider the system of $N_1$ fermions 
(QP's) each with (integral or half-integral) angular momentum $l_1$ 
(i.e., in a LL of degeneracy $g_1=2l_1+1$).
Neglecting the finite-size corrections, this corresponds to the 
filling factor $\nu_1=N_1/g_1$.
Let the fermions form $N_2={1\over2}N_1$ bosonic pairs each with 
angular momentum $l_2=2l_1-\mathcal{R}_1$, where $\mathcal{R}_1$ 
is an odd integer.
The filling factor for the system of pairs, defined as $\nu_2=
N_2/g_2$ where $g_2=2l_2+1$, is given by $\nu_2={1\over4}\nu_1$.
The allowed states of two bosonic pairs are labeled by total 
angular momentum $L=2l_2-\mathcal{R}_2$, where $\mathcal{R}_2$ 
is an even integer.

Of all even values of $\mathcal{R}_2$, the lowest few are not allowed
because of the Pauli exclusion principle applied to the individual 
fermions.
The condition that the two-fermion states with relative angular 
momentum smaller than $\mathcal{R}_1$ are forbidden is equivalent 
to the elimination of the states with $\mathcal{R}_2\le4\mathcal{R}_1$ 
from the two-boson Hilbert space.
Such a ``hard core'' can be accounted for by a CB transformation
with $4\mathcal{R}_1$ magnetic flux quanta attached to each boson.
\cite{theorem}
This procedure defines the effective CB angular momentum $l_2^*=
l_2-2\mathcal{R}_1(N_2-1)$, effective LL degeneracy $g_2^*=g_2-4
\mathcal{R}_1(N_2-1)$, and effective filling factor $\nu_2^*=
(\nu_2^{-1}-4\mathcal{R}_1)^{-1}$.

The CB's defined in this way condense into their only allowed 
$l_2^*=0$ state when the corresponding fermion system has the maximum 
density at which pairing is still possible, $\nu_1=\mathcal{R}_1^{-1}$.
At lower filling factors, the CB-LL is degenerate and the spectrum
of all allowed states of the $N_2$ CB's represents the spectrum of 
the corresponding paired fermion system.
In particular, using the assumption of the superharmonic form of 
boson--boson repulsion, condensed CB states are expected at 
a series of Laughlin filling factors $\nu_2^*=(2q)^{-1}$.
Here, $2q$ is an even integer corresponding to the number of
additional magnetic flux quanta attached to each CB in a subsequent
CB transformation, $l_2^*\rightarrow l_2^{**}=l_2^*-q(N_2-1)$,
to describe Laughlin correlations between the original CB's of
angular momentum $l_2^*$.
From the relation between the fermion and CB filling factors,
$\nu_1^{-1}=(4\nu_2^*)^{-1}+\mathcal{R}_1$, we find the following 
sequence of fractions corresponding to Halperin's pair states, 
$\nu_1^{-1}=q/2+\mathcal{R}_1$.
Finally, we set  $\mathcal{R}_1=1$ for the QE's and $\mathcal{R}_1=3$ 
for the QH's, and use the hierarchy equation,\cite{hierarchy} 
\begin{equation}
\label{eq_hierarchy}
   \nu^{-1}=2p+(1\pm\nu_{\rm QP})^{-1}, 
\end{equation}
to calculate the following sequences of electron filling factors 
$\nu$ derived from the parent $\nu=(2p+1)^{-1}$ state
\begin{equation}
\label{eq_lps}
   \nu^{-1}=2p+1\mp(2+q/2)^{-1}.
\end{equation}
In Eqs.~(\ref{eq_hierarchy}) and (\ref{eq_lps}), the upper sign 
corresponds to the QE's and the lower one to the QH's.
Remarkably, all fractions reported by Pan {\sl et al.} are among 
those predicted for the $\nu={1\over3}$ parent.

The $l_2^{**}=0$ condition for the condensation of the CB's into 
a Laughlin $\nu_2^*=(2q)^{-1}$ state allows the prediction of 
the values of $2l\equiv2l_1$ at which these states should occur
in finite systems of $N\equiv N_1$ QP's.
The result is\cite{fivehalf}
\begin{equation}
\label{eq_seq}
   2l={q+2\over2}N-(q+1).
\end{equation}
Interestingly, this result can be also obtained from the following 
picture.
Let us arrange an even number of particles ($\bullet$) in a shell 
by grouping them into pairs and separating each neighboring pairs 
by a number $q$ of empty states ($\circ$) between them (e.g., 
$\bullet\bullet$ $\circ\circ$ $\bullet\bullet$ $\cdots$ $\circ\circ$ 
$\bullet\bullet$ represents such paired configuration for $q=2$; 
note that the sequence begins and ends with a pair).
Eq.(\ref{eq_seq}) is then obtained by the equation of the total 
number of filled and empty states, ${1\over2}N(q+2)-q$, with the 
angular momentum shell degeneracy, $2l+1$.
The success of this picture is reminiscent of a Laughlin $\nu=
(2p+1)^{-1}$ state that can be pictured as single particles 
separated by $2p$ spaces (e.g., $\bullet$ $\circ\circ$ $\bullet$
$\circ\circ$ $\cdots$ $\bullet$ $\circ\circ$ $\bullet$ to represent 
$\nu={1\over3}$; note that different numbers of spaces correspond 
to an attachment of two flux quanta to a particle and to a pair).

For $q=1$ and 4, Eq.(\ref{eq_seq}) gives $2l={3\over2}N-2$ and $3N-5$, 
respectively.
Note the difference from $2l={3\over2}N$ characteristic of the Jain 
$\nu={2\over3}$ state and and $2l=3N-3$ of the Laughlin $\nu={1\over3}$ 
state.
This difference allows the distinction of Halperin's paired states 
from the Laughlin--Jain states based on the numerical spectra of 
small systems.
On the other hand, $2l=2N-3$ predicted for $q=2$ coincides
\cite{greiter} with the value characteristic of a Moore--Read 
state\cite{moore} describing a half-filled LL$_1$.
The only series of nondegenerate ground states that we found 
numerically in finite systems are at $2l=2N-3$ and $3N-7$ 
(and at their particle--hole conjugate values, $2l=2N+1$ and 
${3\over2}N+2$, obtained by the replacement of $N$ by $2l+1-N$).

\section{Numerical Results}

\subsection{Model}

Our numerical exact diagonalization calculations were carried out 
on Haldane sphere.\cite{haldane83}
In this geometry, $N$ particles are confined in a degenerate shell 
of angular momentum $l$.
The single-particle states are labeled by $m=-l$, $-l+1$, \dots, $l$.
The two-body interaction matrix elements are connected with the 
pseudopotential parameters through the Clebsh-Gordan coefficients. 
The $N$-body interaction Hamiltonian is diagonalized numerically
using a Lanczos algorithm to give the set of low energy states
labeled by total angular momentum $L$.

Standard numerical calculations for $N_e$ electrons are not useful 
for studying the new observed FQH states at $\nu={3\over8}$, 
${4\over11}$, ${5\over13}$, etc., because convincing results require 
values of $N_e$ too large to be diagonalized exactly.
As these states involve pairing of Laughlin QP's and possible Laughlin
correlations between the QP pairs, at least three such pairs must be 
considered.
For $\nu={3\over8}$ this occurs for $N_e=14$ electrons with $2l=33$,
which seems beyond reach of exact diagonalization and explains the 
lack of earlier numerical evidence for incompressibility of this state.
For other states, such as $\nu={4\over11}$, the systems become even 
larger.

Therefore, instead of diagonalizing the $N_e$-electron Hamiltonian, 
we use the QE--QE pseudopotential shown in Fig.~\ref{fig1}(d) and 
diagonalize the (much smaller) interaction Hamiltonian of the 
$N$-QE systems.
This approach is expected to accurately reproduce the low-energy 
spectra of interacting electrons at filling factors $\nu$ between 
${1\over3}$ and ${2\over5}$ (up to an overall constant containing
the energy of the underlying Laughlin $\nu={1\over3}$ state and
the QE creation energies, $\varepsilon_{\rm QE}$ for each QE).
It is justified by fact that the QE--QE interaction energy $V_{\rm QE}$ 
is small compared to the energy gap for creation of additional QE--QH 
pairs, $\varepsilon_{\rm QE}+\varepsilon_{\rm QH}$.
As a result, it is well-known that in this range of $\nu$, 
the low-energy states of (strongly interacting) electrons contain
the (weakly interacting) QE's moving in an underlying (rigid) 
Laughlin $\nu={1\over3}$ fluid.
In the CF picture, this approximation corresponds to neglecting
the inter-LL excitations of CF's and only including the dynamics 
within the partially filled CF-LL$_1$.
In smaller systems, containing up to four QE's or QH's, this 
approximation has been successfully tested by direct comparison 
with the exact $N_e$-electron calculation.\cite{parentage,sitko-fl}
In larger systems, it has recently been used by Lee {\sl et al.}
\cite{lee02}

\begin{figure}
\resizebox{3.4in}{1.73in}{\includegraphics{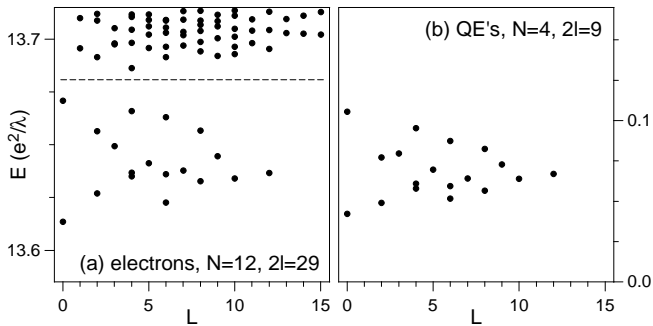}}
\caption{
\label{fig4}
   Energy spectra (energy $E$ as a function of angular momentum 
   $L$) calculated on Haldane sphere for $N=12$ electrons in LL$_0$
   with $2l=29$ (a) and for $N=4$ QE's in CF-LL$_1$ with $2l=9$ (b).
   The energy scale is the same in both frames, but the QE 
   spectrum is only determined up to a constant.}
\end{figure}

Accuracy of this approach is demonstrated in Fig.~\ref{fig4},
where we compare the energy spectra of two systems connected by 
a mean field CF transformation: 
(a) $N=12$ electrons in the LL$_0$ shell with $2l=29$ and 
(b) $N=4$ QE's with $2l=9$.
The four-QE energies, obtained using the pseudopotential of
Fig.~\ref{fig1}(d), are only determined up to an additive
constant, but the structure and relative energies are 
virtually identical in the two spectra.
The agreement can still be noticeably improved by using the QE--QE 
pseudopotential of Fig.~\ref{fig1}(b) obtained for $N=10$ electrons 
(yielding the same $2l=9$ for the pair of QE's).
However, a small residual discrepancy cannot be eliminated by 
fitting $V_{\rm QE}(\mathcal{R})$.
It is due to the fact that (although remarkably accurate) the 
description in terms of pair QE--QE interactions (relying on 
the conservation of QE and QH numbers, i.e., on the lack of 
inter-CF-LL excitations) is not exact.
Note also that using the same pseudopotential parameters 
$V_{\rm QE}(\mathcal{R})$ obtained in large systems\cite{lee01} 
for the calculation of two-body interaction matrix elements 
at different (smaller) values of $2l$ eliminates the finite-size 
effects due to surface curvature, and thus improves accuracy of 
the diagonalization.\cite{he} 

Let us add the following comment about Fig.~\ref{fig4}.
Because $N=12$ electrons at $2l=29$ have an $L=0$ ground state, 
and because the value of $2l=9$ for $N=4$ QE's coincides with
$3N-3$ of a Laughlin $\nu={1\over3}$ state, this single spectrum 
was earlier erroneously interpreted\cite{sitko-cfh,mandal} as 
a success of the CF hierarchy applied to the QE's, and this 
state was incorrectly assigned filling factor $\nu={4\over11}$.
However, upon the analysis of correlations in this state and 
similar spectra of larger systems, it becomes evident that the 
value $2l=9$ must be interpreted at $2N+1$, this four-QE state 
is a particle--hole conjugate of the $2N-3$ sequence, and it 
should be assigned QE and electron filling factors $\nu_{\rm 
QE}={1\over2}$ and $\nu={3\over8}$, respectively.

\subsection{Energy Spectra, Series of Incompressible Ground 
            States, and Excitation Gaps}

We begin with a few examples of the energy spectra of up to 
$N=14$ QE's.
\begin{figure}
\resizebox{3.4in}{3.17in}{\includegraphics{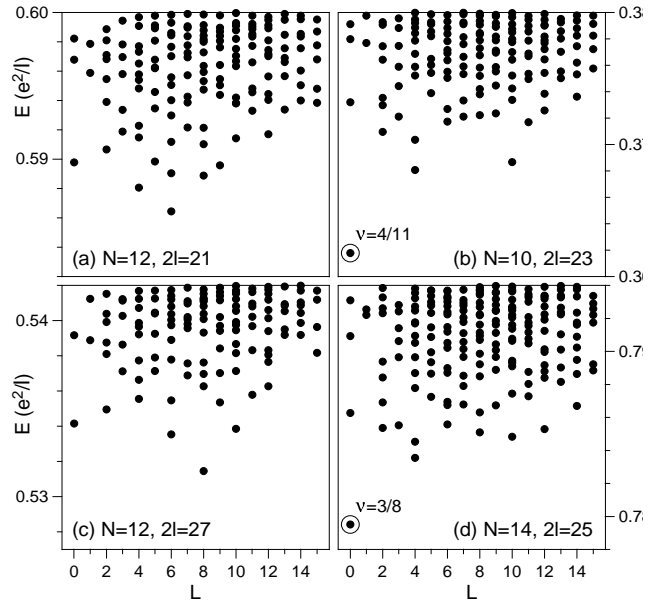}}
\caption{
\label{fig5}
   Energy spectra (energy $E$ as a function of angular momentum 
   $L$) of up to $N=14$ QE's in LL shells with various degeneracies 
   $2l+1$, calculated on Haldane sphere using QE--QE interaction 
   pseudopotential of Fig.~\ref{fig1}(d).}
\end{figure}
Different frames in Fig.~\ref{fig5} show the spectra for: 
$N=12$ and $2l=21$ (a), 
$N=10$ and $2l=23$ (b), 
$N=12$ and $2l=27$ (c), and 
$N=14$ and $2l=25$ (d).
Using the CF picture, these values of $(N,2l)$ can easily be converted 
to $N_e=N+(2l-1)$ and $2l_e=2(l-1)+2(N_e-1)$, characterizing the actual 
electron system (here, $2l-1$ is the degeneracy of the completely filled 
CF-LL$_0$ with angular momentum $l-1$).
Thus, the four $N$-QE systems in Fig.~\ref{fig5} correspond to:
$N_e=32$ and $2l_e=81$ (a), 
$N_e=32$ and $2l_e=83$ (b), 
$N_e=38$ and $2l_e=99$ (c), 
$N_e=38$ and $2l_e=97$ (d).
It is evident that in frames (b) and (d) the FQH-like nondegenerate 
($L=0$) ground states occur, separated from the excited states by 
a gap $\Delta\sim5\cdot10^{-3}$ $e^2/\lambda$.
On the other hand, in frames (a) and (c), the ground state is 
degenerate
($L\ne0$) and no similar gap is observed.
The excitation gaps $\Delta$ in (b) and (d) are larger than other 
energy spacings in these spectra.
This indicates that they are due to the QE--QE interactions rather 
than due to the size quantization in a finite system, and thus that 
they will not vanish in the thermodynamic ($N\rightarrow\infty$) limit.
As we show below, the $L=0$ ground states in Fig.~\ref{fig5}(b) and 
(d) correspond to $\nu={4\over11}$ and ${3\over8}$ in this limit.

We have calculated similar $(N,2l)$ spectra for up to 14 QE's
at filling factors $\nu_{\rm QE}\sim N/(2l+1)$ between ${1\over2}$
and ${1\over3}$. 
Note that the assignment of the filling factor to a finite system
$(N,2l)$ is not trivial and it depends on the form of correlations.
The $(N,2l)$ sequences that correspond to a filling factor $\nu$ 
in the thermodynamic limit are described by a linear relation, 
\begin{equation}
\label{eq_shift}
   2l=N/\nu-\gamma_\nu,
\end{equation}
where the ``shift'' $\gamma_\nu$ depends on the microscopic nature 
of the many-body state causing incompressibility at this $\nu$.
For example, the sequence of finite-size nondegenerate ($L=0$) 
ground states that extrapolates to $\nu={1\over3}$ occurs at $2l=3N-3$ 
for the Laughlin state, at $2l=3N-5$ for the Halperin paired state,
\cite{fivehalf,qepair} and at $2l=3N-7$ for the incompressible QE 
state identified below.

In Tab.~\ref{tab1} we present the excitation gaps obtained for the 
QE systems with various values of $N$ and $2l$.
\begin{table}
\caption{\label{tab1}
   Excitation gaps $\Delta$, in units of $10^{-3}\,e^2/\lambda$, 
   above the nondegenerate ($L=0$) ground states of $N$ QE's each 
   with angular momentum $l$, interacting through pseudopotential 
   in Fig.~\ref{fig1}(d).
   Circles ($\circ$) mark degenerate ($L\ne0$) ground states.
   The values in boldface are the largest; they all belong to the
   three $(N,2l)$ sequences corresponding to $\nu_{\rm QE}={1\over2}$,
   ${1\over3}$, and ${2\over3}$.}
\begin{ruledtabular}
\footnotesize
\begin{tabular}{r|cccccccccccccccc}
 $_N$$^{2l}$ &17&18&19&20&21&22&23&24&25&26&27&28&29\\
\hline
8&{\bf4.71}&$\circ$&$\circ$&$\circ$&0.01&&&&&&&&\\
9&$\circ$&$\circ$&$\circ$&{\bf5.47}&$\circ$&$\circ$&$\circ$
      &1.18&&&&&\\
10&{\bf4.71}&$\circ$&$\circ$&$\circ$&$\circ$&$\circ$
      &{\bf6.29}&$\circ$&0.81&$\circ$&$\circ$&&\\
11&$\circ$&$\circ$&$\circ$&$\circ$&$\circ$&$\circ$&$\circ$
      &$\circ$&$\circ$&{\bf6.07}&$\circ$&$\circ$&$\circ$\\
12&&$\circ$&$\circ$&{\bf5.47}&$\circ$&$\circ$&0.37
      &$\circ$&{\bf4.02}&$\circ$&$\circ$&$\circ$&{\bf5.28}\\
13&&&&$\circ$&$\circ$&$\circ$&$\circ$&$\circ$&$\circ$
      &$\circ$&$\circ$&$\circ$&$\circ$\\
14&&&&&0.01&$\circ$&{\bf6.29}&$\circ$&{\bf4.02}&$\circ$
      &$\circ$&$\circ$&$\circ$\\
15&&&&&&&$\circ$&$\circ$&$\circ$&$\circ$&$\circ$&$\circ$
      &$\circ$\\
16&&&&&&&&1.18&0.81&{\bf6.07}&$\circ$&$\circ$&$\circ$\\
17&&&&&&&&&&$\circ$&$\circ$&$\circ$&$\circ$\\
18&&&&&&&&&&&$\circ$&$\circ$&{\bf5.28}
\end{tabular}
\end{ruledtabular}
\end{table}
The table is symmetric under the replacement of $N$ by 
$2l+1-N$ which reflects the particle--hole symmetry in a 
partially filled QE shell (i.e., in CF-LL$_1$).
This symmetry is only approximate in real systems, but here 
it appears exact because of neglecting the inter-LL 
excitations of the CF's in our model.
The largest of the gaps $\Delta$ (those shown in boldface) 
occur for the following two $(N,2l)$ sequences: $2l=3N-7$
and $2N-3$, corresponding to $\nu_{\rm QE}={1\over3}$ and 
${1\over2}$.
Their particle--hole conjugates series (also in boldface) 
occur at $2l={3\over2}N+2$ and $2N+1$, corresponding to
$\nu_{\rm QE}=1-{1\over3}={2\over3}$ and $1-{1\over2}=
{1\over2}$, respectively.
Using Eq.~(\ref{eq_hierarchy}), these values can be 
converted to the electron filling factors $\nu={3\over8}$, 
${4\over11}$, and ${5\over13}$.

\begin{figure}
\resizebox{3.4in}{1.82in}{\includegraphics{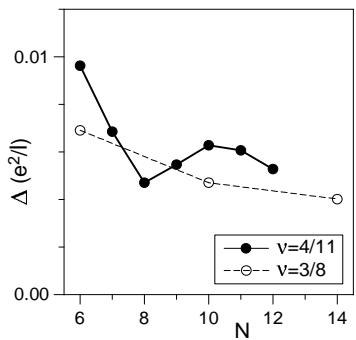}}
\caption{
\label{fig6}
   Excitation gaps $\Delta$ for the $\nu_{\rm QE}={1\over3}$ 
   series of $N$-QE ground states at $2l=3N-7$ (full dots) 
   and for the $\nu_{\rm QE}={1\over2}$ series at $2l=2N-3$ 
   (open circles), plotted as a function of the QE number, $N$.}
\end{figure}

The dependence of the excitation gaps $\Delta$ on the QE 
number $N$ for the $\nu_{\rm QE}={1\over3}$ series at $2l=3N-7$ 
(full dots) and for the $\nu_{\rm QE}={1\over2}$ series at 
$2l=2N-3$ (open circles) is plotted in Fig.~\ref{fig6}.
It is difficult to accurately extrapolate our finite-size data 
to the thermodynamic limit to predict the magnitude of $\Delta$ 
in an infinite (planar) system.
However, we are confident that these two series of finite-size  
nondegenerate ground states describe the FQH states observed 
experimentally at $\nu={4\over11}$ and ${3\over8}$.
The gaps for the larger $N$ are $\Delta\sim5\cdot10^{-3}$ 
$e^2/\lambda$. 
For the experimental situation of Ref.~\onlinecite{pan} (GaAs 
and $B=12.5$~T) this corresponds to $\Delta\sim0.1$~meV or 
$\sim1K$, which seems to be a reasonable value considering the 
fact that the $\nu={4\over11}$ state has only been observed at 
temperatures as low as $T=35$~mK.

\subsection{The $\nu_{\rm QE}={1\over3}$ ($\nu={4\over11}$) State}

The ``shift'' defined by Eq.~(\ref{eq_shift}) and describing the 
$2l=3N-7$ sequence identified here ($\gamma=7$) is different not 
only from $\gamma=3$ describing a Laughlin state, but also from 
$\gamma=5$ that results for Halperin's paired state (with $q=4$).
This precludes the interpretation of these finite-size 
$\nu_{\rm QE}={1\over3}$ ground states found numerically 
(and thus also of the experimentally observed $\nu={4\over11}$ 
FQH state) as either Laughlin or Halperin (paired) state of 
QE's (i.e., particles in the partially filled CF-LL$_1$).
Certainly, the fact that (despite being incompressible) these 
states are not Laughlin states was expected from the fact 
that QE's form pairs over a wide range of $\nu_{\rm QE}
\approx{1\over3}$ (and in the whole low-energy band states, 
not only in the ground states).
However, it is far more surprising that Halperin's paired 
state of QE's turns out as an invalid description for these 
states as well.
Clearly, the correlations between the QE pairs at $\nu_{\rm 
QE}={1\over3}$ must be of a different, non-Laughlin type,
and we do not have an alternative explanation for the 
incompressibility of this state.

This result is consistent with the form of the QE$_2$--QE$_2$ 
pseudopotential shown in Fig.~\ref{fig3}.
Because $V_{{\rm QE}_2}(\mathcal{R}_2)$ is only superharmonic
at $\mathcal{R}_2=6$, the only Laughlin state expected for 
QE$_2$'s could be the one in which all values of 
$\mathcal{R}_2=4$ and 6 are simultaneously avoided.
This possibly valid Halperin's paired state corresponds to 
$q=2$ and $\nu_{\rm QE}={1\over2}$ in Eq.~(\ref{eq_lps}), 
while the $\nu_{\rm QE}={1\over3}$ state corresponds to 
$q=4$ and it would have to avoid all four lowest values of
$\mathcal{R}_2=4$, 6, 8, and 10, which certainly cannot be 
expected from the form of $V_{{\rm QE}_2}(\mathcal{R}_2)$.

While we do not completely understand the correlations between
QE pairs at $\nu_{\rm QE}={1\over3}$, it may be noteworthy that
the value of $\gamma=7$ appropriate for the series of 
incompressible states found here can be obtained for the Laughlin 
state of QE triplets (QE$_3$'s), each with the maximum allowed 
angular momentum, $L=3l-3$.
Such state would be pictured as $\bullet\bullet\bullet$ 
$\circ\circ\circ$ $\circ\circ\circ$ $\bullet\bullet\bullet$
$\cdots$ $\circ\circ\circ$ $\circ\circ\circ$ $\bullet\bullet
\bullet$ with each two closest QE triplets separated by six 
vacancies. 
The idea of particles grouping into triplets or larger clusters 
has been studied in more detail by Read and Rezayi\cite{read}
in the context of electrons in LL$_1$.
Although we do not yet have enough evidence for such particular 
grouping of QE's, let alone for Laughlin correlations between 
the clusters, this possibility is definitely worth further 
investigation, especially for the predicted exotic (parafermion) 
statistics of the excitations of such hypothetical ground state.
\cite{read}
Note, however, that the numerical results show an $L=0$ ground 
state at $2l=3N-7$ for every integral value of $N$, which seems 
inconsistent with the idea of complete clustering of QE's into 
molecules of any size.
It can also be noticed that partial pairing with ${1\over3}N$ of 
QE pairs and ${1\over3}N$ of unpaired QE's also leads to $2l=3N-7$, 
but again, only for values of $N$ that are divisible by three.

\subsection{The $\nu_{\rm QE}={1\over2}$ ($\nu={3\over8}$) State}

The other sequence of finite-size $L=0$ ground states identified
in Tab.~\ref{tab1} occurs at $2l=2N-3$, i.e., at the same value as
for the Moore--Read states of electrons half-filling LL$_1$.
\cite{moore,greiter,morf}
This value also coincides\cite{fivehalf} with the value predicted 
for Halperin's paired state with $q=2$, in which the eigenstates 
of two QE$_2$ bosonic pairs corresponding to the two lowest values 
of $\mathcal{R}_2=4$ and 6, are avoided.
Because of the subharmonic behavior of $V_{{\rm QE}_2}(\mathcal{R}_2)$ 
at $\mathcal{R}_2\ne6$ (see Fig.~\ref{fig3}), this $q=2$ state is 
the only Halperin paired state of the series given by 
Eq.~(\ref{eq_lps}) that might possibly occur in a QE system.

However, despite the facts that this sequence occurs at the 
predicted value of $2l=2N-3$ and only for even numbers of QE's
(as expected for paired states), its interpretation as a Halperin
paired state (or Moore--Read state) of QE's turns out incorrect.
First indication is that it only seems to occur for odd numbers 
of QE pairs, ${1\over2}N=3$, 5, and 7, while the ground states 
for ${1\over2}N=4$ and 6 (at $2l=13$ and 21, respectively) both 
turn out degenerate.
Unfortunately, we do not have data for ${1\over2}N>7$ to confirm 
our expectation that the finite-size ground states at $2l=2N-3$ 
have $L=0$ and a large excitation gap for all odd values of 
${1\over2}N$.
Note also that the state found here for $N=10$ and $2l=17$ happens 
to be a particle--hole conjugate state of $N=8$ QE's at the same 
value $2l$ (i.e., it belongs to the $2l={3\over2}N+2$ sequence), 
and thus we only find two $L=0$ ground states ($N=6$ and 14) that 
are unique for the $2N-3$ series. 

More direct proof for the $\nu_{\rm QE}={1\over2}$ state not being 
Halperin's paired state (or a related Moore--Read state) comes 
from the analysis of its three-body correlations.\cite{3body} 
We find significant occupation of the compact triplet state QE$_3$ 
with the minimum allowed relative angular momentum $\mathcal{T}=3$
at $\nu_{\rm QE}={1\over2}$, which is inconsistent with the picture
of Laughlin-correlated (i.e., spatially separated) pairs.
This is in contrast with the behavior of the Moore--Read paired 
state (an exact trial state that describes Halperin-like pairing
at a half-filling) that is characterized by having exactly zero 
occupation of the $\mathcal{T}=3$ triplet state.\cite{greiter}
In fact, we have calculated squared overlaps $\zeta$ of the 
finite-size $\nu_{\rm QE}={1\over2}$ states with the Moore--Read 
states of QE's and they turn out {\em very small} (e.g., $\zeta=
0.03$ for $N=14$) and insensitive to the parity of ${1\over2}N$.
Nevertheless, despite the fact that we do not yet understand the 
correlations in the $\nu_{\rm QE}={1\over2}$ state (e.g., the 
importance of ${1\over2}N$ being odd in finite systems), we believe 
that the $2l=2N-3$ series identified here indeed describes the 
observed $\nu={3\over8}$ FQH state.

\section{Results for model interactions}

In this section we present the results of similar calculations, 
obtained using a model pseudopotential $U_\alpha(\mathcal{R})$ 
instead of $V_{\rm QE}(\mathcal{R})$.
Its only non-vanishing coefficients are 
\begin{eqnarray}
\label{eq_u}
   U_\alpha(1)&=&1-\alpha,\nonumber\\
   U_\alpha(3)&=&\alpha/2.
\end{eqnarray}
It is known\cite{fivehalf} that the correlations characteristic 
of electrons in the partially filled LL$_0$ and LL$_1$ are 
accurately reproduced by $U_\alpha$ with $\alpha\approx0$ 
and ${1\over2}$, respectively.
Similarly, by the comparison of pair amplitudes, we have confirmed 
that $U_\alpha$ with $\alpha\approx1$ causes correlations 
characteristic of QE's in their partially filled LL.

We have repeated the diagonalization of a few finite systems 
with $2l=2N-3$ and $3N-7$, for $\alpha$ varying between 0 and 1, 
in order to answer the following two questions.
First, to what extent is the stability of the identified 
$\nu={3\over8}$ and ${4\over11}$ states affected by the 
(width-dependent) details of the QE--QE interaction?
And second, does a phase transition occur for values of 
$\alpha$ between ${1\over2}$ and 1, indicating a different 
origin of the incompressibility of the $\nu={3\over8}$ and 
${4\over11}$ states and their electron counterparts 
(in LL$_1$) at $\nu={5\over2}$ and ${7\over3}$?
The latter question is naturally motivated by our two 
observations: 
(i) the $2l=2N-3$ sequence of nondegenerate ground states 
occurs only for odd numbers of QE pairs (${1\over2}N=3$, 5, 
and 7), in contrast to the situation in LL$_1$ where they 
occurred for any value of ${1\over2}N$, and (ii) the 
$\nu_{\rm QE}={1\over2}$ has small overlap with the 
Moore--Read state (of QE's).

\begin{figure}
\resizebox{3.4in}{1.82in}{\includegraphics{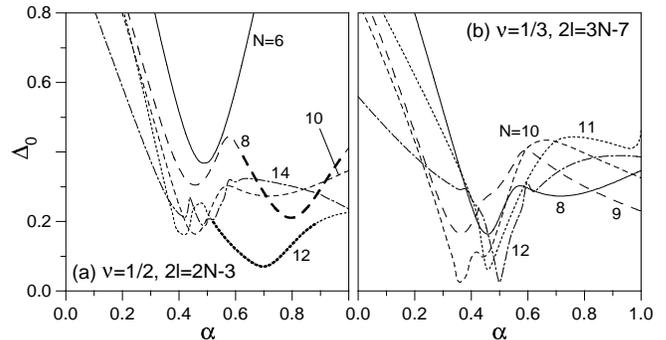}}
\caption{
\label{fig7}
   The excitation gap $\Delta_0$ between the lowest and the 
   first excited states in the $L=0$ subspace of $N$ particles 
   on Haldane sphere with the values of $2l$ corresponding to
   $\nu={1\over2}$ (a) and $\nu={1\over3}$ (b), plotted as 
   a function of the interaction parameter $\alpha$ defined
   by Eq.~(\ref{eq_u}).}
\end{figure}

In Fig.~\ref{fig7} we plot the $L=0$ excitation energy gap 
$\Delta_0$ (difference between the two lowest energy levels 
at $L=0$), as a function of $\alpha$.
A minimum in $\Delta_0(\alpha)$ signals a (forbidden) level 
crossing, i.e., a phase transition in the $L=0$ subspace.
Such minima occur near $\alpha={1\over2}$ for all values of 
$N$ and for both $2l=2N-3$ and $3N-7$.
They reveal destruction of Laughlin correlations that occur
for small $\alpha$ (e.g., for electrons in LL$_0$) and 
formation of incompressible $\nu={1\over2}$ and ${1\over3}$ 
states of a different (paired) character that occur for 
$\alpha\approx{1\over2}$ (e.g., for electrons in LL$_1$).

In Fig.~\ref{fig7}(a), similar strong minima occur at $\alpha
\approx0.7$ for $N=8$ and 12 (marked with thick lines).
This is consistent with our observation that the correlations 
between the QE's and between the electrons in LL$_1$ (both at 
the half-filling) are different.
In Fig.~\ref{fig7}(a) and (b), additional weaker minima between 
$\alpha={1\over2}$ and 1 appear also for other combinations of 
$N$ and $2l$.
This confirms that the $\nu={1\over2}$ and ${1\over3}$ incompressible 
states of QE's are generally different from those of the electrons 
in LL$_1$, despite the fact that they both usually occur at the 
same values of $2l=2N-3$ and $3N-7$ in the finite systems.

\begin{figure}
\resizebox{3.4in}{1.82in}{\includegraphics{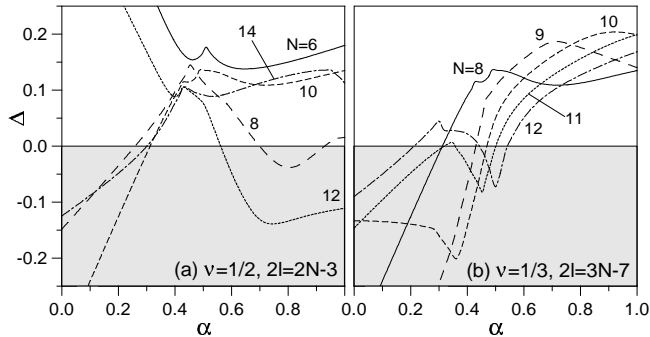}}
\caption{
\label{fig8}
   The excitation gap $\Delta$ from the lowest state with $L=0$ 
   to the lowest state with $L\ne0$ for $N$ particles on Haldane 
   sphere with the values of $2l$ corresponding to $\nu={1\over2}$ 
   (a) and $\nu={1\over3}$ (b), plotted as a function of the 
   interaction parameter $\alpha$ defined by Eq.~(\ref{eq_u}).}
\end{figure}

The absolute excitation gaps $\Delta(\alpha)$ of the $L=0$ ground 
states (difference between the lowest energies at $L\ne0$ and $L=0$) 
are shown in Fig.~\ref{fig8}.
The negative value of $\Delta$ means that the absolute ground state
is degenerate (i.e., has $L\ne0$), and the abrupt changes in the 
slope of $\Delta(\alpha)$ occur whenever level crossings occur for 
the lowest $L\ne0$ state.
Clearly, except for $N=8$ and 12 with $2l=2N-3$, the lowest $L=0$ 
states remain the absolute ground states of the system in the whole
range of $\alpha$ between ${1\over2}$ and 1.
This was first noticed by Greiter {\sl et al.}\cite{greiter} for $N=10$
at half-filling, and it implies that the incompressibility of the 
$\nu_{\rm QE}={1\over2}$ and ${1\over3}$ ground states will not 
be easily destroyed in experimental systems by a minor deviation 
from the model QE--QE pseudopotential used here in the numerical 
diagonalization.

\begin{figure}
\resizebox{3.4in}{2.39in}{\includegraphics{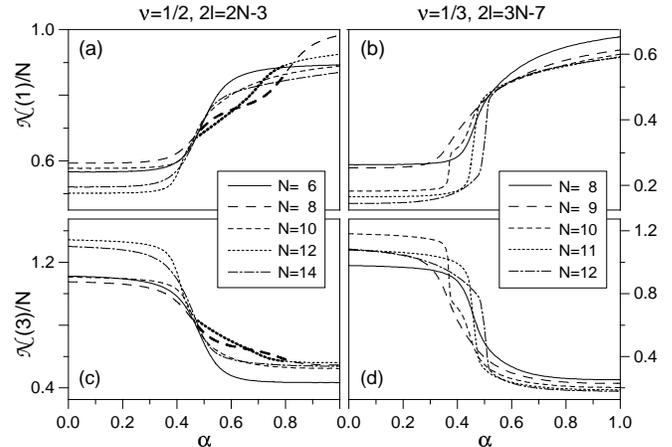}}
\caption{
\label{fig9}
   The average number of pairs with relative angular momentum 
   $\mathcal{R}=1$ (a,b) and $\mathcal{R}=3$ (c,d) per particle, 
   $\mathcal{N}(\mathcal{R})/N$, calculated for the lowest state 
   in the $L=0$ subspace of $N$ particles on Haldane sphere with 
   the values of $2l$ corresponding to $\nu={1\over2}$ (a,c) and 
   $\nu={1\over3}$ (b,d), plotted as a function of the interaction 
   parameter $\alpha$ defined by Eq.~(\ref{eq_u}).}
\end{figure}

Let us finally examine the dependence of the leading pair amplitudes, 
$\mathcal{G}(1)$ and $\mathcal{G}(3)$, on $\alpha$.
In Fig.~\ref{fig9} we plot the number of pairs $\mathcal{N}
(\mathcal{R})={1\over2}N(N-1)\mathcal{G}(\mathcal{R})$, 
divided by $N$.
A transition from Laughlin correlations at $\alpha\sim0$ to pairing 
at $\alpha\sim{1\over2}$ (and possibly grouping into larger clusters
at $\alpha\sim1$) is clearly visible in each curve.
It is also confirmed that just as the Laughlin ground state remains 
virtually insensitive to the exact form of the interaction 
pseudopotential $V_e$ as long as it is strongly {\em superharmonic} 
at short range, the correlations in the $\nu_{\rm QE}={1\over2}$ and 
states ${1\over3}$ are quite independent of the details of the QE--QE 
interaction, as long as $V_{\rm QE}$ is strongly {\em subharmonic} at 
short range.
This result supports our expectation that the incompressible QE ground 
states found here numerically indeed describe the FQH $\nu={3\over8}$ 
and ${4\over11}$ electron states observed in experiment.

On the other hand, correlations at $\alpha\approx{1\over2}$ 
(electrons in LL$_1$), characterized by having $\mathcal{G}(1)
\approx\mathcal{G}(3)$, are quite different from those at 
$\alpha\sim1$ (QE's), characterized by having the minimum possible 
$\mathcal{G}(3)$, much smaller than $\mathcal{G}(1)$.
Finally, with thick lines in Fig.~\ref{fig9}(a) we have marked the 
curves for $N=8$ and 12 in the vicinity of $\alpha\sim0.7$ at which 
the forbidden crossings were found in Fig.~\ref{fig7}(a).
A different behavior of $\mathcal{N}(1)/N$ and $\mathcal{N}(3)/N$ 
for these two values of $N$ is clearly visible.

\section{Conclusions}

Using exact numerical diagonalization in Haldane spherical geometry, 
we have studied the energy spectra and wavefunctions of up to $N=14$ 
interacting QE's in the Laughlin $\nu={1\over3}$ parent state (i.e., 
CF's each carrying two flux quanta).
We have demonstrated by direct calculation of the pair amplitudes
$\mathcal{G}(\mathcal{R})$ that, at their sufficiently large filling 
factor ($\nu_{\rm QE}>{1\over5}$), the QE's form pairs or larger 
clusters, with a significant occupation of the minimum relative 
pair angular momentum, $\mathcal{R}=1$.
The QE (and analogous QH) clustering is an opposite behavior to 
Laughlin correlations characterizing e.g.\ electrons partially 
filling LL$_0$.
Therefore it invalidates the reapplication of the CF picture to the 
individual QE's or QH's (and thus also the equivalent multi-flavor 
CF model) and precludes the simple hierarchy interpretation of any 
incompressible states at ${6\over17}<\nu<{2\over5}$ or ${2\over7}<
\nu<{6\over19}$.

The series of finite-size nondegenerate ground states at QE filling 
factors $\nu_{\rm QE}={1\over2}$, ${1\over3}$, and ${2\over3}$ have 
been identified.
These values correspond to the electronic filling factors $\nu=
{3\over8}$, ${4\over11}$, and ${5\over13}$, at which the FQH effect 
has recently been discovered.\cite{pan}
Due to a discussed similarity between the QE--QE and QH--QH 
interactions, these three QE states have their QH counterparts at 
$\nu_{\rm QH}={1\over4}$, ${1\over5}$, and ${2\over7}$, corresponding 
to $\nu={3\over10}$, ${4\over13}$, and ${5\over17}$, all of which have 
also been experimentally observed.\cite{pan}
Finally, it is argued that the reported\cite{pan} $\nu={6\over17}$ 
FQH state might be a standard hierarchy state (Laughlin $\nu_{\rm QE}
={1\over5}$ state), although it could only be observed in sufficiently
wide systems.
Its QH counterpart at $\nu={6\over19}$ (Laughlin $\nu_{\rm QH}={1\over7}$ 
state) would require a larger width than $\nu={6\over17}$ which might 
explain why it has not yet been observed.

The finite-size $\nu_{\rm QE}={1\over2}$, ${1\over3}$, and ${2\over3}$ 
states of QE's (CF's in LL$_1$) are found at the same values of $2l=
2N-3$, $3N-7$, and ${3\over2}N+2$ as the $\nu={5\over2}$ (Moore--Read
\cite{moore,greiter,morf}), ${7\over3}$, and ${8\over3}$ FQH states 
of electrons in LL$_1$, respectively, despite the different electron 
and CF pseudopotentials.
Therefore we have studied the dependence of the wavefunctions and 
stability of the novel FQH states on the exact form of interaction 
at short range.
We found several indications that the novel QE states are distinctly 
different from the electron states in LL$_1$: 
(i) the $\nu_{\rm QE}={1\over2}$ state appears incompressible only 
for the odd values of ${1\over2}N$;
(ii) the pair-correlation functions $\mathcal{G}(\mathcal{R})$ 
(and, especially, the triplet-correlation functions\cite{3body})
are quite different;
(iii) although they remain incompressible, the ground states 
appear to undergo phase transitions when the QE--QE pseudopotential 
is continuously transformed into that of electrons in LL$_1$;
(iv) the overlaps with the electron states in LL$_1$ and with the 
Moore--Read trial state are very small.
However, further studies are needed to understand these transitions.
On the other hand, weak dependence of the wavefunctions and 
excitation gaps of the novel FQH states on the details of the 
QE--QE interaction (as long as it remains strongly subharmonic at 
short range) justifies the use of a model pseudopotential in 
the realistic numerical calculation.

We have also explored Halperin's idea\cite{halperin83,qepair} of 
the formation of Laughlin states of QE pairs (QE$_2$'s).
An appropriate composite boson model has been formulated and shown
to predict a family of novel FQH states at a series of fractions 
including all those observed in experiment.
However, several observations strongly point against this simple 
model: (i) the QE$_2$--QE$_2$ interaction pseudopotential is not 
superharmonic to support Laughlin correlations of QE$_2$'s 
(except possibly for $\nu_{\rm QE}={1\over2}$);
(ii) the values of $2l$ predicted for finite $N$ are different 
from these obtained from the numerical diagonalization
(except for $\nu_{\rm QE}={1\over2}$);
(iii) the numerical results do not confirm the significance 
of parity of the number of QE's in finite systems (the 
$\nu_{\rm QE}={1\over2}$ states occur only for $N=6$, 10, 
and 14, and the $\nu_{\rm QE}={1\over3}$ states occur for 
both even and odd values of $N$);
(iv) the analysis of three-body correlations suggests formation
of clusters larger than pairs.\cite{3body}
In fact, despite an earlier expectation,\cite{greiter} we find 
\cite{3body} that Halperin's pairing idea is far more appropriate 
for the electrons in LL$_1$ than for QE's in CF-LL$_1$.

We have not found evidence for only partial pairing (and possibly 
Laughlin-correlated mixed states of pairs and unpaired electrons) 
or grouping of QE's into larger clusters of well-defined size
(and possibly Laughlin correlations between them\cite{read}).
However, further investigation of both these ideas is necessary.
Also, since the experiment\cite{pan} indicates complete spin 
polarization of the novel FQH states, here we have not studied 
unpolarized systems, considered in great detail in a number of 
earlier studies begun with the work of Park and Jain.\cite{park}
Finally, the connection between the QE pairing studied here and
recent shot-noise experiments\cite{comforti} indicating bunching 
of QP's in Laughlin and Jain FQH states at ultra-low temperatures
is not yet clear.

\acknowledgments

The authors thank V. J. Goldman, M. Shayegan, and R. G. Mani 
for helpful discussions.
This work was supported by Grant DE-FG 02-97ER45657 of the Materials
Science Program -- Basic Energy Sciences of the U.S. Dept.\ of Energy.  
AW acknowledges support from Grant 2P03B02424 of the Polish KBN.
KSY acknowledges support from Grant R14-2002-029-01002-0 of the KOSEF.

\end{document}